\documentclass[11pt]{article}
\usepackage{moriond,amsmath}
\def\be{\begin{equation}}
\def\ee{\end{equation}}
\def\bea{\begin{eqnarray}}
\def\eea{\end{eqnarray}}

\def\d{{\rm d}}
\def\bsp#1\esp{\begin{split}#1\end{split}}

\newcommand{\Photo}{\includegraphics[height=35mm]{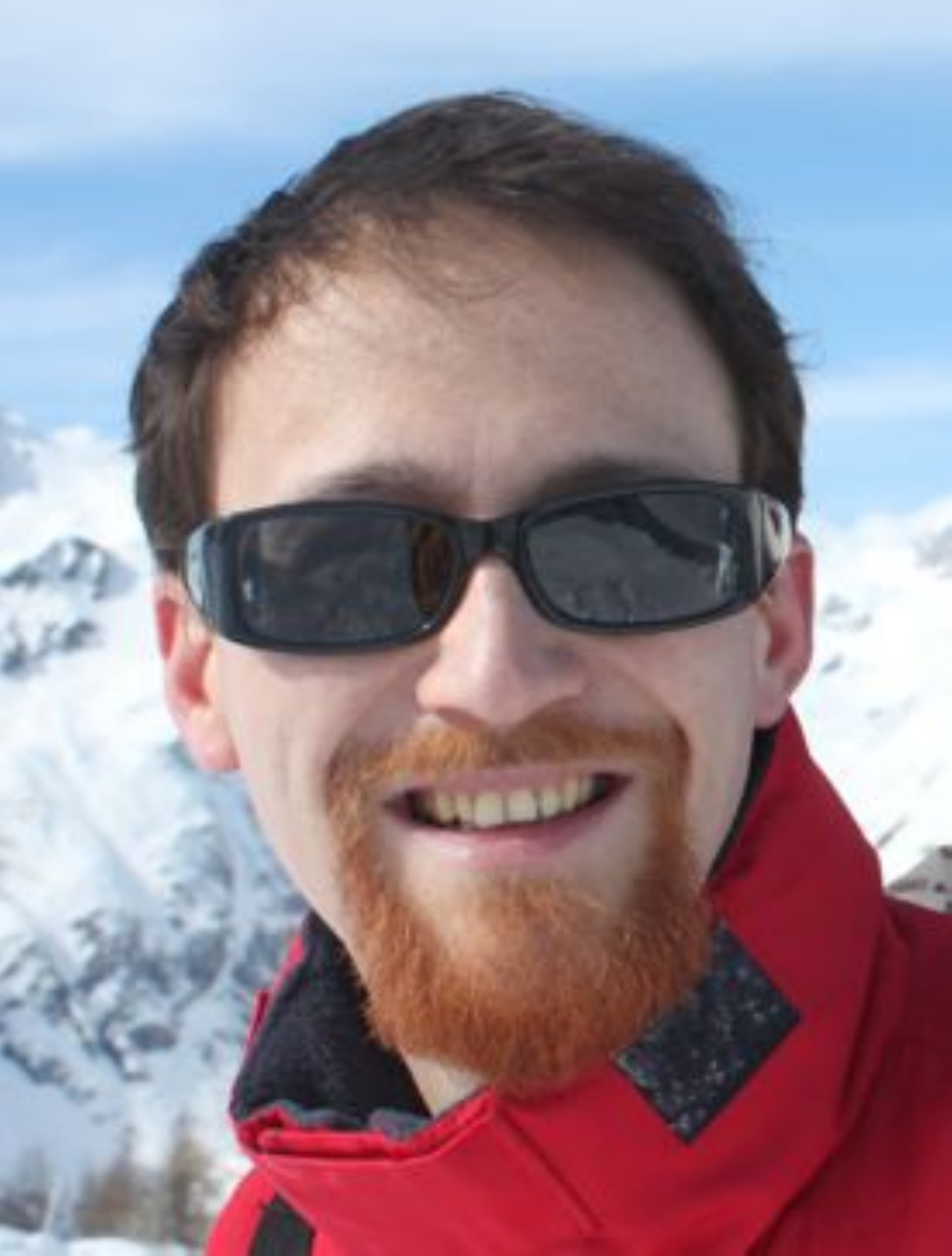}}

\begin{document}
\vspace*{4cm}
\title{QCD RESUMMATION IN THE FRAMEWORK OF SUPERSYMMETRY}

\author{BENJAMIN FUKS$^{1,2}$ \footnote{Speaker.},
MICHAEL KLASEN$^3$, DAVID R.\ LAMPREA$^3$, MARCEL ROTHERING$^3$}
\address{
  $^1$Theory Division, Physics Department, CERN, CH-1211 Geneva 23, Switzerland\\
  $^2$Institut Pluridisciplinaire Hubert Curien/D\'epartement Recherches Subatomiques,\\ 
    Universit\'e de Strasbourg/CNRS-IN2P3, 23 Rue du Loess, F-67037 Strasbourg, France\\
  $^3$Institut f\"ur Theoretische Physik, Westf\"alische Wilhelms-Universit\"at M\"unster,\\
   Wilhelm-Klemm-Stra\ss{}e 9, D-48149 M\"unster, Germany}

\maketitle\abstracts{
  Motivated by current searches for electroweak superpartners at the Large Hadron Collider,
  we present precision predictions for pair production of such particles
  in the framework of the Minimal Supersymmetric Standard Model. We make use of various QCD 
  resummation formalisms and
  match the results to pure perturbative QCD computations. We study the impact of scale variations
  and compare our results to predictions obtained by means of
  traditionally used  Monte Carlo event generators.}

\section{Introduction}
\begin{flushright}
\vspace{-17cm}{\small CERN-PH-TH-2013-096, IPHC-PHENO-13-04, MS-TP-13-11}
\vspace{16.5cm}
\end{flushright}
After almost half a century of theoretical developments and experimental discoveries in
high-energy physics, an extremely coherent picture arises as the so-called Standard Model of
particle physics. Since this theory contains a fundamental scalar field, the
stabilization of its mass with respect to radiative corrections is questionable. This has led to
a plethora of new physics models among which
weak-scale supersymmetry \cite{Nilles:1983ge,Haber:1984rc} (SUSY) is one of the most
appealing option since it encompasses in addition, \textit{e.g.},
gauge coupling unification and a candidate for dark
matter.

The current non-observation of any hint for strong superpartners has
shifted the experimental attention to
the production of electroweak sleptons, neutralinos and charginos.
Investigations at the Large Hadron Collider (LHC),
at a center-of-mass energy of $\sqrt{S_h}=$ 7 and 8~TeV,
have already allowed to impose
bounds of several hundreds of GeV on their masses~\cite{ATLAS:2012uks,CMS:aro}. These analyses
however rely on leading order (LO) computations \cite{Bozzi:2004qq,Bozzi:2007me,%
Debove:2008nr} supplemented by QCD next-to-leading order (NLO) corrections
\cite{Beenakker:1999xh}. Since such predictions suffer
from rather large theoretical uncertainties,
soft-gluon resummation of the large logarithmic terms arising at small transverse momentum
or close to the production threshold have to be accounted for and matched
to fixed order \cite{Bozzi:2006fw,Bozzi:2007qr,%
Bozzi:2007tea,Debove:2009ia,Debove:2010kf,Debove:2011xj,Fuks:2012qx}.

We briefly review, in Section \ref{sec:theo}, three resummation
formalisms that can be employed for such precision calculations and illustrate their
main effects in Section \ref{sec:results} for gaugino pair
production. In addition, we also confront
the resummed predictions to results obtained using
LO Monte Carlo event generators including multiparton matrix
element merging after parton showering. We summarize our work in Section \ref{sec:conclusions}.

\section{Soft gluon resummation: a brief insight} \label{sec:theo}
We focus on the hadroproduction of pairs of electroweak superpartners with an
invariant mass $M$ and a transverse momentum $p_T$.
After a Mellin transform with respect to $M^2/S_h$, the differential
cross section $\d^2\sigma/\d M^2 \d p_T^2$
can be expressed, in  conjugate $N$-space, as a product of the
partonic cross section $\sigma_{ab}$ with the
densities $f_{a,b}$ of the partons $a,b$ in the colliding hadrons,
\be
  M^2 \frac{\d^2\sigma}{\d M^2 \d p_T^2}(N-1)=\sum_{ab} f_a(N,\mu_F^2)
    f_b(N,\mu_F^2) \sigma_{ab}(N,M^2,p_T^2,\mu_F^2, \mu_R^2)\ .
\ee
Under this form where factorization and renormalization
scales $\mu_F$ and $\mu_R$ are explicitly indicated,
we can resum to all orders in the strong coupling $\alpha_s$ the large
logarithmic terms arising when $p_T$ tends
towards zero and/or close to the production
threshold. In this case, the partonic cross section can be refactorized into a closed
exponential form, respectively reading 
\bea
 \sigma_{ab}^{\rm (res.)}(N,M^2,\mu_F^2, \mu_R^2) =
  {\cal H}_{ab}(M^2, \mu_F^2,\mu_R^2) \exp\Big[{\cal G}_{ab}(N,M^2,\mu_F^2,\mu_R^2)\Big] \ , \\
 \sigma_{ab}^{\rm (res.)}(N,M^2,p_T^2,\mu_F^2, \mu^2_R)\! = \!
   \int_0^\infty \d b\frac{b}{2}\ J_0(bp_T)\
  {\cal H}_{ab}(M^2, \mu_F^2,\mu_R^2) \exp\Big[{\cal G}_{ab}(N,b,M^2,\mu_F^2,\mu_R^2)\Big] \ ,
\label{eq:res2}\eea
in the threshold (after integrating upon $p_T$) and small-$p_T$ regime.
The hard part of the cross section is described by the function ${\cal H}_{ab}$ whereas the
Sudakov form factor ${\cal G}_{ab}$ embeds
soft and collinear parton radiation and absorbs the large logarithms.
Eq.~\eqref{eq:res2} also contains an inverse Fourier transform,
$J_0$ denoting the $0^{\rm th}$-order Bessel function, so that the singularities
of the integrand have to be handled after deforming the integration contour
into the complex plane \cite{Laenen:2000de}.

Although the logarithmic contributions must be resummed when they are large,
the full perturbative computation, only partially accounted for by resummation, is expected to be
reliable otherwise. Therefore, the
fixed order ($\sigma^{\rm (f.o.)}$) and resummed ($\sigma^{\rm (res.)}$) results have to
be consistently combined by subtracting from
their sum their overlap $\sigma^{\rm (exp.)}$,
\be
 \sigma_{ab}=
 \sigma^{\rm(res.)}_{ab}+\sigma^{\rm(f.o.)}_{ab}-\sigma^{\rm(exp.)}_{ab} \ .
\ee
Since both $\sigma^{\rm(res.)}$ and $\sigma^{\rm (exp.)}$
are computed in Mellin space, an inverse transform is in order.
To handle the singularities arising at the level of the $N$-space
cross section, the integration contour is distorted following the principal value
procedure and minimal prescription~\cite{Contopanagos:1993yq,Catani:1996yz}.

The form of the quantities introduced above depends on the resummation regime.
Transverse-momentum resummation deals with logarithms arising
at small $p_T$, while threshold resummation takes care of those appearing close to the
production threshold. Finally, joint resummation allows for resumming both types of logarithms
simultaneously. We refer to the {\sc Resummino} manual and references therein for the
relevant analytical expressions at the next-to-leading logarithmic (NLL)
accuracy~\cite{Fuks:2013vua}.

\section{Gaugino pair production at the next-to-leading logarithmic accuracy}\label{sec:results}

\begin{figure}
\begin{minipage}{0.42\linewidth}
\centerline{\includegraphics[width=0.9\linewidth]{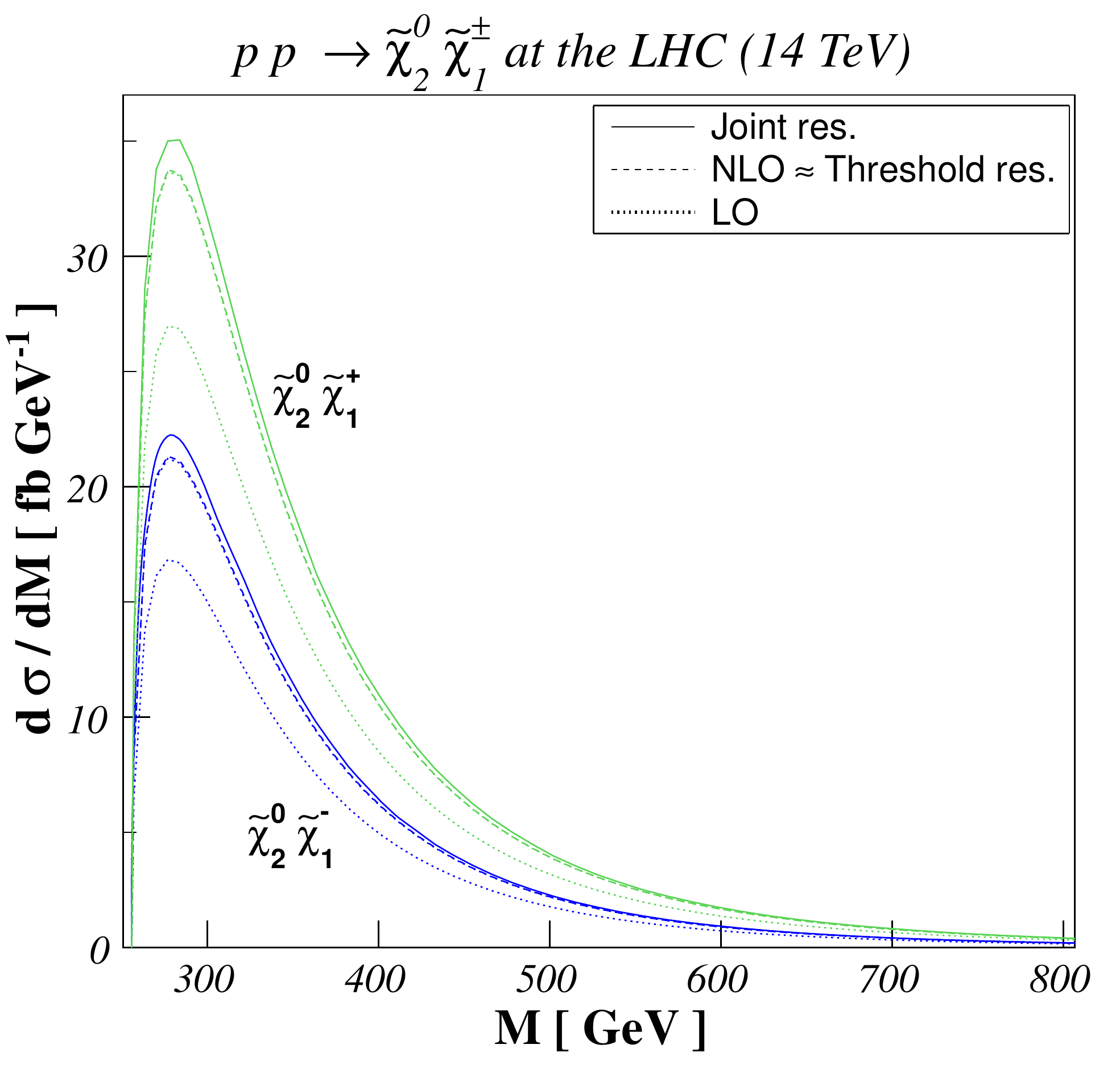}}
\end{minipage}
\hfill
\begin{minipage}{0.42\linewidth}
\centerline{\includegraphics[width=0.9\linewidth]{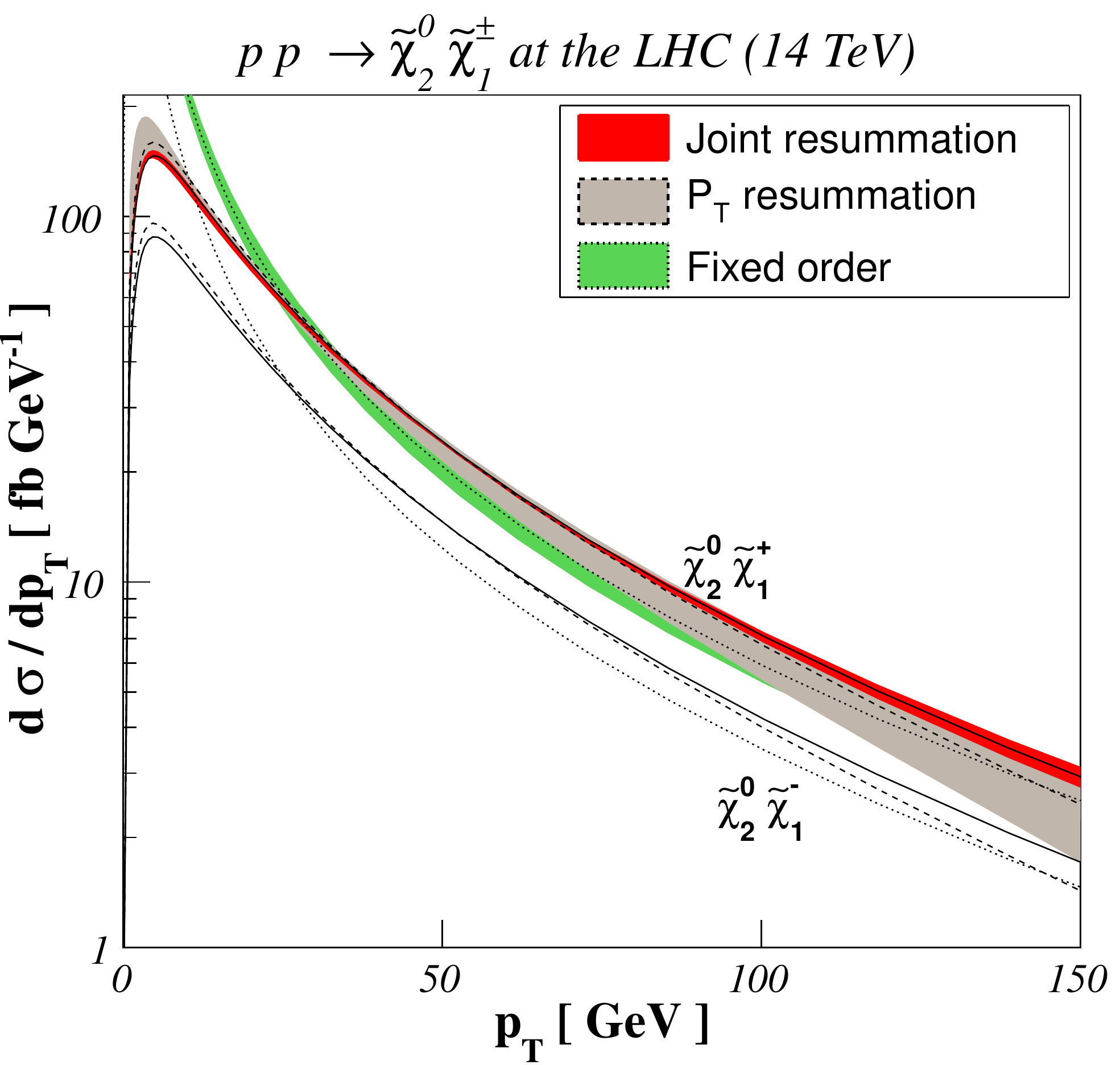}}
\end{minipage}
\caption{Invariant mass (left) and transverse-momentum (right) distributions of
an associated $\tilde\chi_2^0\tilde\chi_1^\pm$ pair
produced at the LHC, at fixed order and after matching to resummation. Scale uncertainties are
indicated for the $p_T$ spectra.}
\label{fig:1}
\end{figure}

In Fig.~\ref{fig:1}, we address the production of an associated $\tilde\chi^\pm_1
\tilde \chi_2^0$ pair at the LHC, for $\sqrt{S_h}= 14$~TeV. We adopt the
LM9  benchmark scenario \cite{Ball:2007zza}, where both gauginos have a mass of about
150 GeV whereas gluinos and squarks lie above 1 TeV, and employ
the CTEQ6 parton densities \cite{Nadolsky:2008zw}. On the left panel of the figure,
we present spectra in the invariant mass of the gaugino pair. The LO results (dotted)
are found to be considerably
smaller than NLO predictions with or without matching to NLL resummation.
Since we restrict the distribution to the small invariant-mass region, far from
the production threshold, threshold resummation
does not lead to a significant effect with respect to NLO (dashed). In contrast,
jointly resummed predictions (full) exceed the NLO ones due to the resummation
of the large logarithms arising at small $p_T$.

On the right panel of Fig.~\ref{fig:1}, we show transverse-momentum
spectra of the gaugino pair. While fixed-order predictions
at ${\cal O}(\alpha_s)$ (dotted) diverge at small
$p_T$ due to uncanceled soft singularities from real gluon emission,
resummed calculations exhibit a pronounced peak.
For intermediate values of $p_T$, resummation effects are found to be still important
with a $K$-factor greater than unity.
We finally show that calculations using $p_T$ (dashed) and joint (full) resummation agree
with each other, although the scale uncertainty associated with the latter, estimated by
varying both unphysical scales by a factor of two around the
average mass of the two gauginos, is considerably smaller due to the resummation of threshold
logarithms.

\begin{figure}
\begin{minipage}{0.46\linewidth}
\centerline{\includegraphics[width=0.9\linewidth]{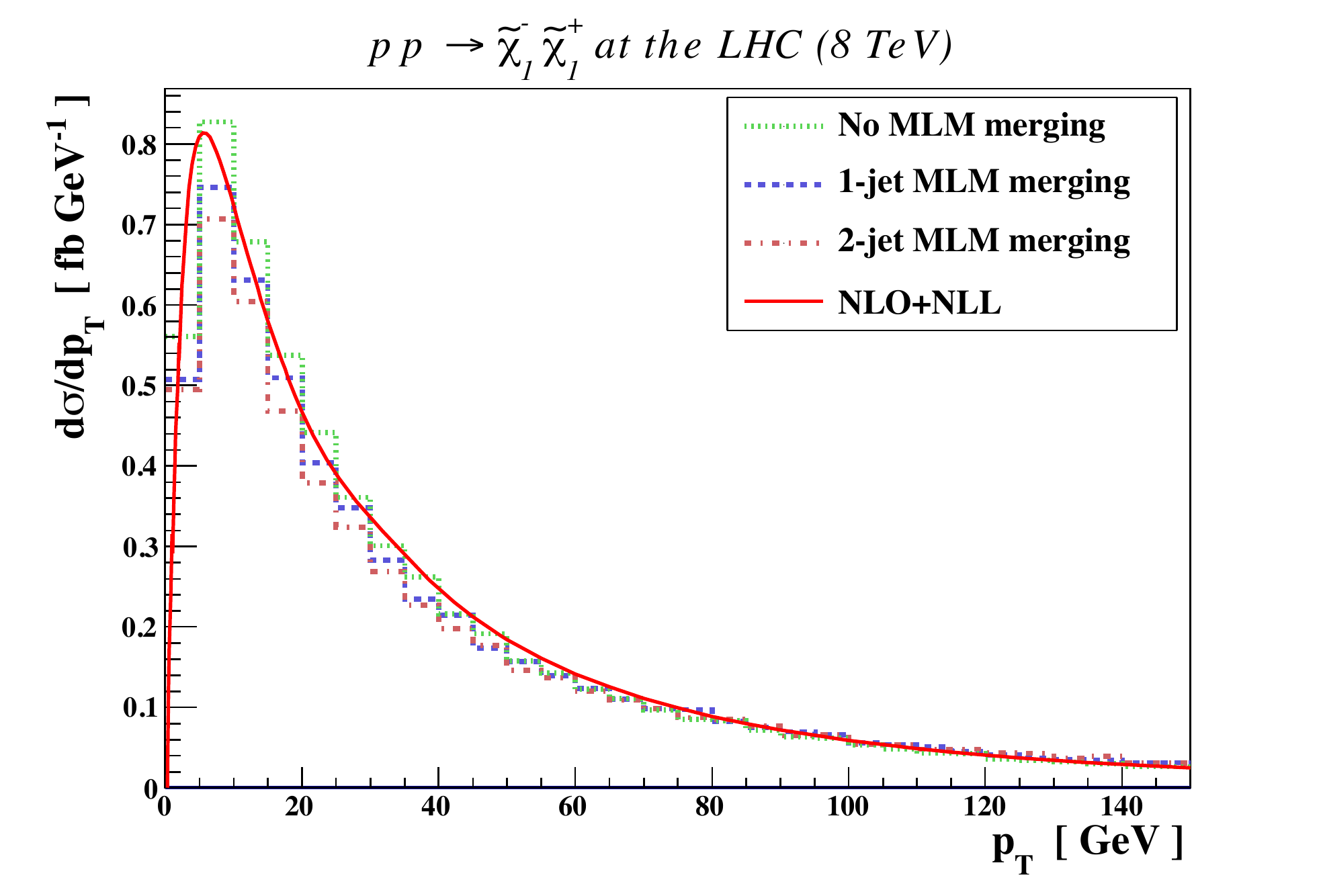}}
\end{minipage}
\hfill
\begin{minipage}{0.46\linewidth}
\centerline{\includegraphics[width=0.9\linewidth]{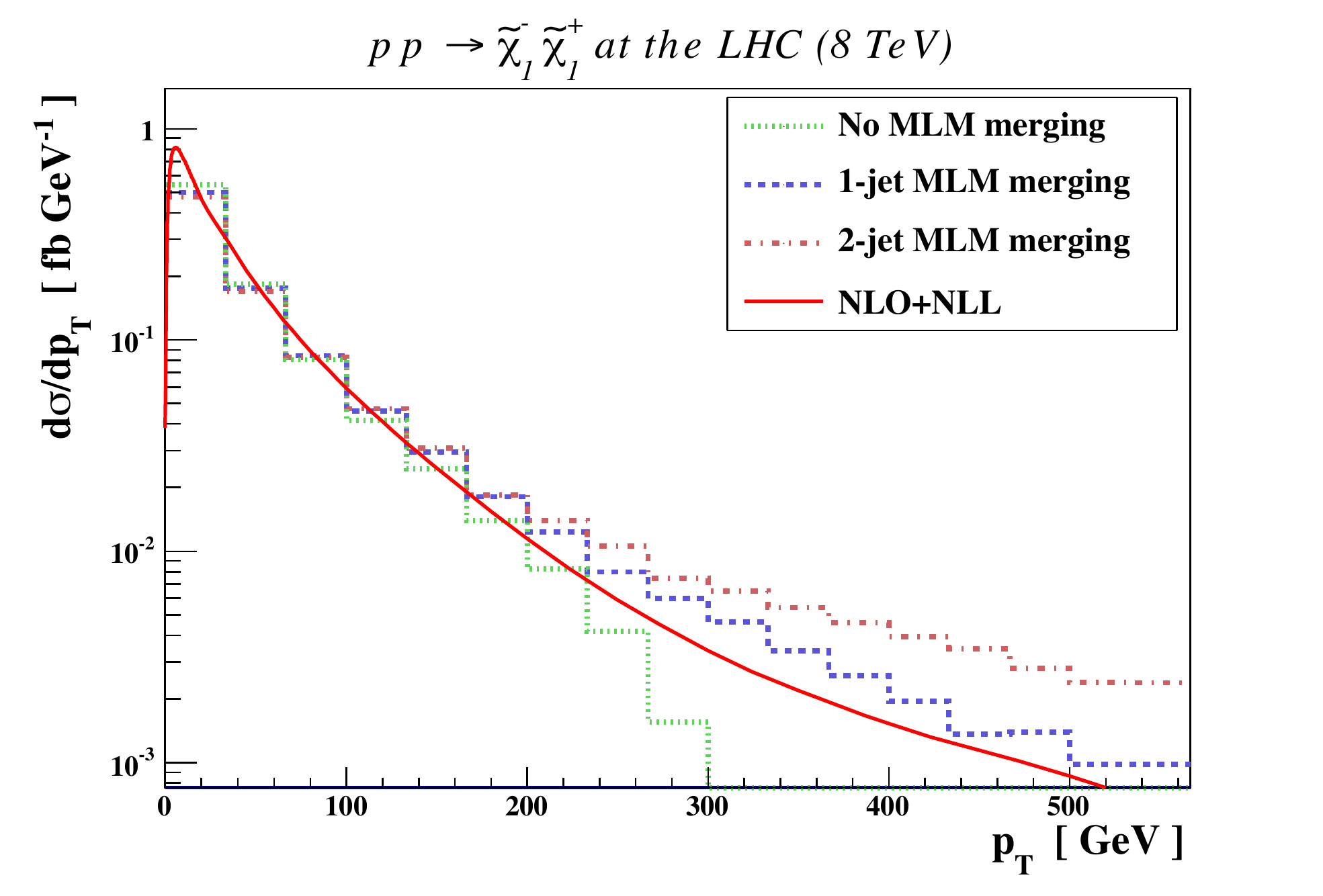}}
\end{minipage}
\caption{Distributions in the transverse momentum of a
  $\tilde\chi^+_1\tilde\chi^-_1$ pair produced at the LHC. We compare resummation results
  to several approaches by means of LO event generators including MLM merging
  techniques.}
\label{fig:2}
\end{figure}

In Fig.~\ref{fig:2}, we focus on $\tilde\chi^+_1\tilde\chi^-_1$ production
at $\sqrt{S_h} \!=\! 8$~TeV and on the first
SUSY scenario proposed by the LPCC \cite{AbdusSalam:2011fc}. It
embeds sub-TeV squarks and gluinos and a lightest chargino of about 300 GeV.
Using {\sc MadAnalysis}~5~\cite{Conte:2012fm}, we confront joint resummation (full) to
predictions of the LO event generator {\sc MadGraph}~5~\cite{Alwall:2011uj},
matched to  {\sc Pythia} 6 \cite{Sjostrand:2006za} for parton showering, the
necessary UFO module \cite{Degrande:2011ua} being exported from
{\sc FeynRules} \cite{Christensen:2008py,Christensen:2009jx,Christensen:2010wz,%
Duhr:2011se,Fuks:2012im}.
We allow the generated events to contain zero (dotted), up to one (dashed)
or up to two (dot-dashed) additional jets
and merge them following the MLM merging scheme \cite{Mangano:2006rw}.
After normalizing the Monte Carlo results to the resummed prediction of 40.51 fb
and employing the MSTW parton densities \cite{Martin:2009iq},
we observe a very good agreement
between all approaches in the small-$p_T$ region. In contrast, 
in the large-$p_T$ region,
only Monte Carlo predictions including up to one extra parton
agree with the resummed results, since both rely on the same matrix elements.

\section{Summary} \label{sec:conclusions}
We have analyzed predictions for electroweak superpartner production at
the LHC obtained by means of different resummation methods after a combination with
NLO predictions. The results have been found to be more reliable and exhibit smaller
uncertainties stemming from scale variation. A similar accuracy can be reached by means
of LO Monte Carlo event generators after merging matrix elements possibly
containing additional partons.

\section*{Acknowledgments}
We acknowledge support from the BMBF
Theorie-Verbund and the Theory-LHC-France initiative of the CNRS/IN2P3.
B.F.~thanks the organizers for the great conference and the invitation.

\end{document}